\documentclass[iop]{emulateapj}
\usepackage{natbib}
\usepackage{comment}
\usepackage{amsmath}
\usepackage{hyperref,breakurl}

\shorttitle{A High-Frequency Radio Background Excess from {\it Planck} }
\shortauthors{MURPHY \& CHARY}
\slugcomment{Accepted to ApJ; May 2, 2018}

\begin{document}
\title{Excess in the High Frequency Radio Background: Insights from {\it Planck}}

\author{Eric J. Murphy}
\affil{National Radio Astronomy Observatory, 520 Edgemont Road, Charlottesville, VA 22903, USA; \email{emurphy@nrao.edu}}
\affil{Infrared Processing and Analysis Center, California Institute of Technology, MC 314-6, Pasadena, CA 91125, USA}
\author{Ranga-Ram Chary}
\affil{Infrared Processing and Analysis Center, California Institute of Technology, MC 314-6, Pasadena, CA 91125, USA;  \email{rchary@caltech.edu}}

\begin{abstract}
We conduct a stacking analysis using 1.4\,GHz NRAO VLA Sky Survey (NVSS) detections and {\it Planck} all-sky maps to estimate the differential source counts down to the few 100\,$\mu$Jy level at 30, 44, 70 and 100\,GHz.  Consequently, we are able to measure the integrated extragalactic background light from discrete sources at these frequencies.   By integrating down to a 1.4\,GHz flux density of $\approx$2$\,\mu$Jy, we measure integrated, extragalactic brightness temperatures from discrete sources of $105.63\pm10.56\,$mK, $21.76\pm3.09\,\mu$K, $8.80\pm0.95\,\mu$K,  $2.59\pm0.27\,\mu$K, and $1.15\pm0.10\,\mu$k at 1.4, 30, 44, 70, and 100\,GHz, respectively.  Our measurement at 1.4\,GHz is slightly larger than previous measurements, most likely due to using NVSS data compared to older interferometric data in the literature, but still remains a factor of $\approx$4.5 below that required to account for the excess extragalactic sky brightness measured at 1.4\,GHz by ARCADE\,2.    The fit to ARCADE\,2 total extragalactic sky brightness measurements is also a factor of $\approx$8.6, 6.6, 6.2, and 4.9 times brighter than what we estimate from discrete sources at 30, 44, 70 and 100\,GHz, respectively.  The extragalactic sky spectrum (i.e., $T_{\rm b} \propto \nu^{\beta}$) from discrete sources appears to flatten with increasing frequency, having a spectral index of $\beta=-2.82\pm0.06$ between 1.4 and 30\,GHz and $\beta=-2.39\pm0.12$ between 30 and 100\,GHz.  We believe that the spectral flattening most likely arises from a combination of Gigahertz-peaked sources and the spectral hardening of radio-detected sources at higher frequencies, particularly at faint flux densities.   However, the precise origin of a hard component of energetic electrons responsible for the emission remains unclear.

\end{abstract}

\section{Introduction}
Our knowledge of the extragalactic radio source population at frequencies spanning $10 \lesssim \nu \lesssim 100\,$GHz is currently poor.  
This largely stems from the fact that wide-field, high-frequency radio surveys are extremely time-consuming using present facilities.  
The combination of primary beam and sensitivity considerations when conducting radio surveys, given that extragalactic radio sources typically have steep spectra (i.e., $S_{\nu} \propto \nu^{\alpha}$, where $\alpha \sim -0.7$ for star-forming galaxies), makes it much more efficient to achieve large samples over wide patches of sky at lower frequencies.   
Consequently, our knowledge on extragalactic radio sources comes from surveys at frequencies of $\sim$GHz and below \citep[e.g.,][]{cm84, jc84b, raw85,gdz10}.  


\begin{figure*}
\epsscale{1.}
\plotone{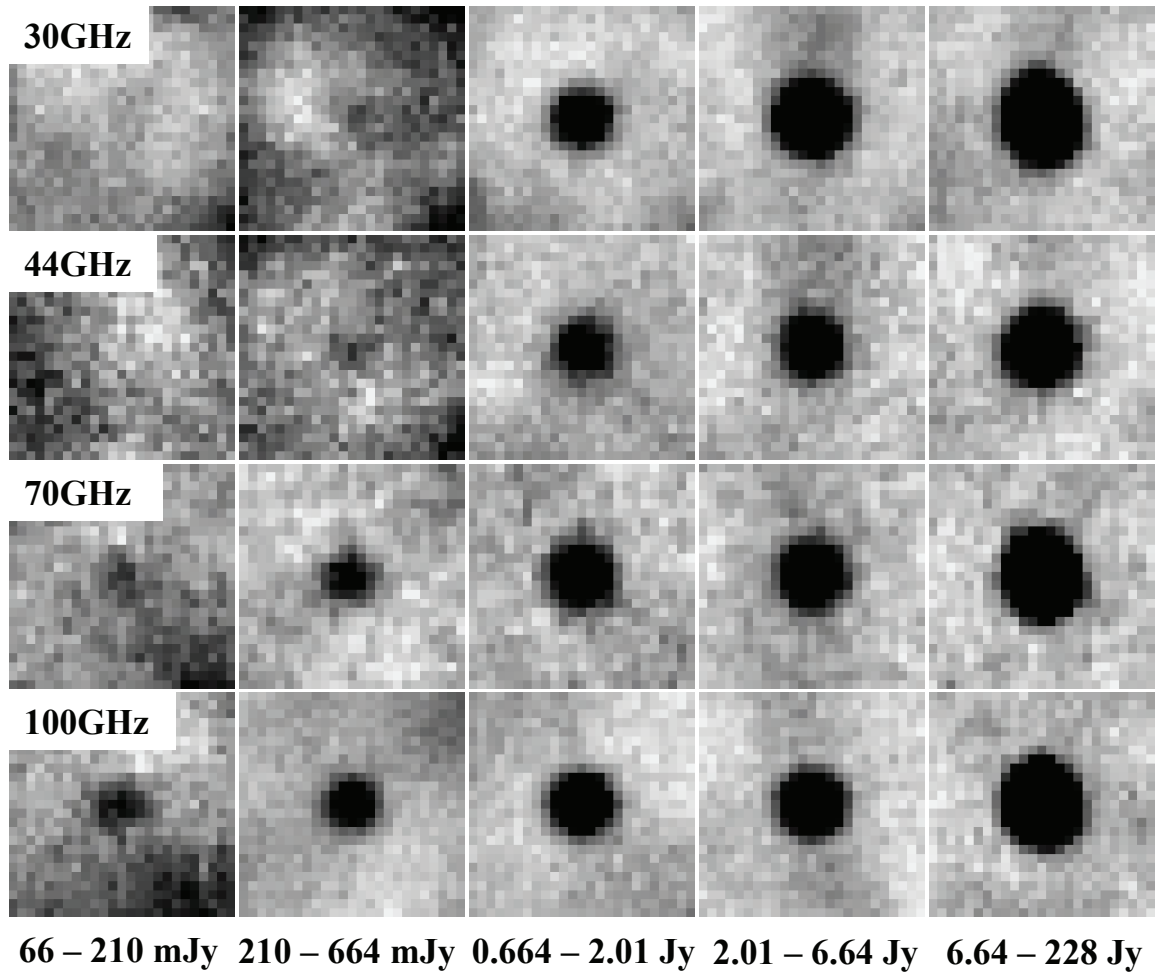}
\caption{The (mean) stacked images for each 1.4\,GHz flux density bin.  
The size of each cutout is 5 times the FWHM of the {\it Planck} beam at the corresponding frequency on a side.  
To illustrate what the stacked images look like when we get into faint 1.4\,GHz flux density bins, we show stacked images for which we were not able to achieve a $>5\,\sigma$ detection.  
We did not obtain a statistically significant detection at either 30 or 44\,GHz in the lowest two 1.4\,GHz flux density bins.  
At 70 and 100\,GHz, we did not obtain a statistically significant detection in the faintest 1.4\,GHz flux density bin shown (see Table \ref{tbl-1}). }
\label{fig:stkimgs}
\end{figure*}

\begin{figure}
\epsscale{1.2}
\plotone{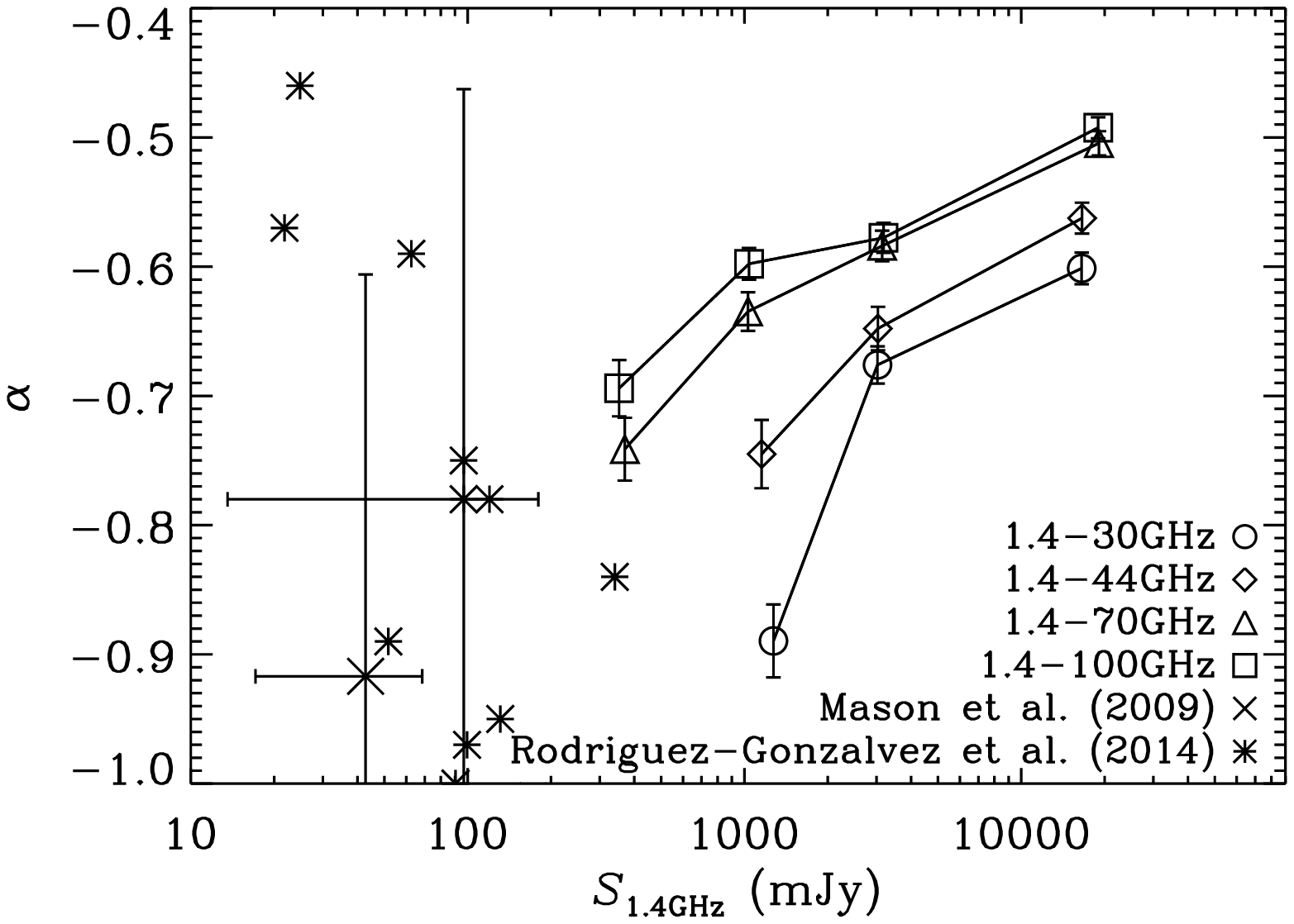}
\caption{The 1.4-to-30, 44, 70, and 100\,GHz spectral indices at each stacked flux density bin in which a $5\,\sigma$ detection was achieved.  
For each frequency, there is a clear trend indicating that the fainter sources typically have steeper spectral indices.  
The 1.4-to-30\,GHz spectral indices given in \citet{crg15} are over plotted; 
the error bars indicate the median and standard deviation, respectively, of their sample.  
We additionally show the average 1.4-to-31\,GHz spectral index reported by \citet{bm09}, where the horizontal error bar indicates the spread in 1.4\,GHz flux densities over which their number counts were estimated in Figure \ref{fig:dnds}, and the vertical error bar illustrates the standard deviation of the mean spectral index in their analysis.   
While standard deviations are shown for \citet{crg15} and  \citet{bm09} as error bars, and are thus not directly comparable to the errors on the mean show for the spectral indices from our stacking analysis, this is done to illustrate that our stacked results are well within measurements from direct detections.  
}
\label{fig:stkspx}
\end{figure}

At such frequencies (i.e., $\nu \lesssim 10$\,GHz), the radio sky is dominated by synchrotron and free-free emission, both of Galactic and extragalactic origins.  
After removing the estimated
contributions of Galactic foreground and the cosmic microwave background (CMB), the Absolute Radiometer for Cosmology, Astrophysics, and Diffuse Emission \citep[ARCADE\,2;][]{djf11} has measured the sky brightness temperature spectrum (i.e., $T_{\rm b} \propto \nu^{-2}S_{\nu} \propto \nu^{\beta}$, where $\beta = \alpha -2$), and report on a significant excess of low-frequency radio emission between 22\,MHz and 10\,GHz.  
At 1.4\,GHz, they estimate an excess brightness temperature of $T_{\rm b} \approx 480$\,mK, which is nearly a factor of 5 larger than the contribution from all known populations of extragalactic sources, and does not seem to be easily explained by even fainter populations of galaxies \citep[see][]{jc12, tv14}.  
Assessing the significance of this excess at higher frequencies is difficult owing to the lack of sensitive, wide-field surveys at frequencies $\gtrsim 10$\,GHz.  

While there has been a significant increase in the number of wide-area surveys at frequencies $\gtrsim$10\,GHz, those with flux density limits of $\sim$10\,mJy typically only cover a few $\sim$10s\,deg$^{2}$ patches of sky \citep[e.g.,][]{act01, emw03, emw10}.  
There are a couple of exceptions that cover over 20,000\,deg$^{2}$, including the {\it Wilkinson Microwave Anisotropy Probe} \citep[\textit{WMAP};][]{wmap} and the Australia Telescope 20\,GHz Survey \citep[AT20G;][]{at20}.  
While {\it WMAP} observed the entire sky at 23, 33, 41, 61, and 94\,GHz, its flux density limit is only $\approx$1\,Jy \citep{elw09}. 
Similarly, AT20G covers $\sim$20,000\,deg$^{2}$, but reaches a flux density limit of $\approx$40\,mJy.  
This is still not as sensitive as lower-frequency GHz surveys such as the NRAO VLA Sky Survey \citep[NVSS;][]{nvss}, which has a flux density limit of $\approx$2.1\,mJy at 1.4\,GHz.  
However, with new {\it Planck} all-sky maps, the combination of all-sky coverage with modest depths at these frequencies, provides the opportunity to characterize the contribution of discrete sources to the extragalactic background at frequencies spanning $30-100$\,GHz. 

Using {\it Planck} all-sky maps, in combination with radio source counts from the NVSS, we use a stacking analysis to estimate the differential source counts at radio frequencies down to the few 100\,$\mu$Jy flux density level, along with the total sky brightness at these frequencies arising from discrete radio sources.  
This paper is organized as follows.  
The data are presented in Section 2, along with a description of the stacking analysis.  
Our results are then presented in Section 3, and discussed in Section 4.  
Finally, in Section 5, we summarize our main conclusions.

\begin{deluxetable*}{c|cccc|cccc}
\tablecaption{Stacking Results \label{tbl-1}}
\tabletypesize{\scriptsize}
\tablewidth{0pt}
\tablehead{
\colhead{1.4\,GHz Flux Density Bin}  & 
\colhead{$S_{\rm 30\,GHz}^{\rm stack}$} & \colhead{$S_{\rm 44\,GHz}^{\rm stack}$} & \colhead{$S_{\rm 70\,GHz}^{\rm stack}$} & \colhead{$S_{\rm 100\,GHz}^{\rm stack}$} &
\colhead{} & \colhead{} & \colhead{} & \colhead{}\\
\colhead{(mJy)} & 
\colhead{(mJy)} & \colhead{(mJy)} & \colhead{(mJy)} & \colhead{(mJy)} &
\colhead{$\alpha^{\rm 30\,GHz}_{\rm 1.4\,GHz}$} & \colhead{$\alpha^{\rm 44\,GHz}_{\rm 1.4\,GHz}$} & \colhead{$\alpha^{\rm 70\,GHz}_{\rm 1.4\,GHz}$} & \colhead{$\alpha^{\rm 100\,GHz}_{\rm 1.4\,GHz}$}
}
\startdata
66$\,\leq\, S_{\rm 1.4\,GHz}\,<\,$210&\nodata&\nodata&6$\,\pm\,$2.4&
4$\,\pm\,$1.3&\nodata&\nodata&\nodata&\nodata\\
210$\,\leq\, S_{\rm 1.4\,GHz}\,<\,$664&4$\,\pm\,$11.4&6$\,\pm\,$8.8&
20$\,\pm\,$1.8&18$\,\pm\,$1.6&\nodata&\nodata&-0.74$\,\pm\,$0.02&
-0.69$\,\pm\,$0.02\\
664$\,\leq\, S_{\rm 1.4\,GHz}\,<\,$2100&87$\,\pm\,$7.0&88$\,\pm\,$7.6&
86$\,\pm\,$4.3&81$\,\pm\,$3.4&-0.89$\,\pm\,$0.03&-0.75$\,\pm\,$0.03&
-0.63$\,\pm\,$0.01&-0.60$\,\pm\,$0.01\\
2100$\,\leq\, S_{\rm 1.4\,GHz}\,<\,$6642&394$\,\pm\,$12.3&324$\,\pm\,$16.2&
319$\,\pm\,$11.4&270$\,\pm\,$10.6&-0.68$\,\pm\,$0.01&-0.65$\,\pm\,$0.02&
-0.58$\,\pm\,$0.01&-0.58$\,\pm\,$0.01\\
6642$\,\leq\, S_{\rm 1.4\,GHz}\,<\,$228260&2699$\,\pm\,$58.0&2382$\,\pm\,$67.4&
2645$\,\pm\,$55.2&2315$\,\pm\,$41.0&-0.60$\,\pm\,$0.01&-0.56$\,\pm\,$0.01&
-0.50$\,\pm\,$0.01&-0.49$\,\pm\,$0.01
\enddata
\tablecomments{The measured central frequencies of the {\it Planck} data, which were used for the analysis, are 28.5, 44.1, 70.3, and 100\,GHz.  Spectral indices were calculated and used in the present analysis only when the signifcance of the stacked {\it Planck} flux densities were $>5\,\sigma$.}
\end{deluxetable*}

\section{Data and Analysis}
The present analysis makes use of {\it integrated} 1.4\,GHz flux densities included in the NVSS component catalogue available on {\it Vizier}\footnote{http://vizier.u-strasbg.fr.  This version of the catalog was constructed from the following deconvolved component catalog created on 2002 September 27, and provided by the NVSS Catalog's authors here: ftp://ftp.cv.nrao.edu/nvss/CATALOG/NVSSCatalog.text.gz}.    
The flux density limit of the NVSS catalog is 2.1\,mJy.  
NVSS sources are only considered outside of a Galactic latitude cut of $\vert b \vert  > 20\degr$.  
The latitude cut is used to mitigate any potential contamination from Galactic sources.  
For strong sources having flux densities larger than $\gtrsim 100$\,Jy, NVSS VLA observations are known to saturate.  
There is one such source in the NVSS sample after making the above mentioned Galactic latitude cut, Vir\,A (= 3C\,274\,S), for which we adopt $S_{\rm 1.4\,GHz} = 226$\,Jy \citep{cb88}. 

We also use the nominal mission 
30, 44, 70\,GHz {\it Planck}/LFI and 100\,GHz {\it Planck}/HFI total intensity maps included in Data Release 1 \citep{Planck2013-I}.  
For simplicity, we refer to the {\it Planck} data by their nominal frequencies, however, for all calculations requiring the frequency of the {\it Planck} data, we use the measured central frequencies of 28.5, 44.1, 70.3, and 100\,GHz.  
Each map is corrected for the dipole signal produced by the combination of the motions of the spacecraft, the Earth, and the Solar System with respect to the CMB.  
Details on the {\it Planck}/LFI calibration and map making can be found in \citep{Planck2013-V}, while details on {\it Planck}/HFI calibration and map making can be found in \citet{Planck2013-VIII}.  
The calibration uncertainty for the {\it Planck} maps is 0.8\% at 30\,GHz, 0.6\% at 44 and 70\,GHz, and 0.4\% at 100\,GHz.

\subsection{Stacking Analysis}
\label{sec-stack}
We created cutout images for all 4 {\it Planck} bands at the location of NVSS sources.  
Each cutout is $25\times25$ pixels, with a pixel size that is 1/5 the FWHM of the {\it Planck} beam at the corresponding frequency.  
Accordingly, the size of each cutout is 5 times the FWHM of the {\it Planck} beam on a side.  
Our stacking analysis was carried out in a way to avoid multiple inclusion of sources in the image stacks.  
This is done by stacking on the location of NVSS sources with decreasing 1.4\,GHz flux density, and removing any other NVSS sources that fall within a radius of three times the FWHM of the {\it Planck} beam at the current position for further consideration.  
The {\it Planck} beam FWHM values at 30, 44, 70, and 100\,GHz are taken as 32\farcm65, 27\farcm00, 13\farcm01, and  9\farcm94, respectively.  
Thus, any occurrence of double counting or inclusion of flux density from bright nearby sources is eliminated.  

Once the cutouts are generated for each {\it Planck} band, we then stack all sources within a 1.4\,GHz flux density bin having a width of 0.5\,dex, except for the brightest bin, for which we allowed the size to be large enough to include $\gtrsim$30 sources.  
The final 1.4\,GHz flux density bins are given in Table \ref{tbl-1}.  
The all-sky {\it Planck} maps at these frequencies are provided in units of thermodynamic temperature (i.e., K$_{\rm CMB}$), requiring a conversion to obtain integrated flux densities from the stacked {\it Planck} cutouts that are in the same units as the integrated 1.4\,GHz flux densities in the NVSS catalog.  
We therefore first convert from units of K$_{\rm CMB}$ to Raleigh-Jeans brightness temperature using the following multiplicative factors of 0.979328, 0.95121302, 0.88140690, and 0.76581996 
at 30, 44, 70, and 100\,GHz, respectively (note, for 1.4\,GHz this value is unity).  
These values were then converted into units of MJy\,sr$^{-1}$ using the multiplicative factors of 24.845597, 59.666236, 151.73238, and 306.81118 at 30, 44, 70, and 100\,GHz, respectively.  

Stacking is done by first removing sky values from the individual cutouts, taken as the median pixel value in an annulus defined by radii of 1 and 2 times the FWHM of the {\it Planck} beams.  
The sky subtracted cutouts are then combined by taking a mean of the pixels at each location in the stack.  
We take the mean stack, as opposed to the median, given that we are working with image cutouts that include detected sources (i.e., not residual maps, for which the latter would be more appropriate).  
The stacked images at 30, 44, 70, and 100\,GHz are shown in Figure \ref{fig:stkimgs}, for which we show images to just below the 1.4\,GHz flux density bins that resulted in a statistically significant detection in the {\it Planck} bands.  
The integrated flux density of the stacked images, in units of mJy, is then taken by summing pixels within an aperture having a radius equal to the FWHM of the {\it Planck} beam, and then multiplying that number by an aperture correction factor of $\approx$1.07.  

Given the significantly larger size of the {\it Planck} beam compared to the 45\arcsec~NVSS beam, we account for multiple NVSS sources that fall within the integration aperture at each frequency by keeping track of their individual 1.4\,GHz flux densities.  
This is done since each NVSS source may be contributing to the stacked {\it Planck} flux density at some level.  
We therefore mimic what {\it Planck} would see at 1.4\,GHz given the angular resolution at each frequency by convolving each NVSS source within a radius of 2.5 times the FWHM of the corresponding {\it Planck} beam using a Gaussian with the same FWHM.  
We then sum the contribution of 1.4\,GHz emission from all sources within the integration aperture (i.e., with a radius equal to the FWHM of the {\it Planck} beam) as the total 1.4\,GHz flux density associated with the {\it Planck} source.  
At a distance of 2.5 times the FWHM of the beam, a source contributes at the $\approx$0.1\% level to the total flux density measured within our integration aperture.    

The uncertainty of the stacked flux density is estimated through a Monte Carlo exercise.  
The above described stacking procedure was carried out 100 times on each {\it Planck} map, at a random position located at a distance that was between 1 and 4 times the FWHM of the corresponding {\it Planck} beam from the center of the NVSS position.  
The standard deviation of the {\it Planck} flux densities at the random positions was then taken as an estimate on the uncertainty for the stacked flux density in that 1.4\,GHz flux density bin.  
The stacked flux densities, along with estimates on their uncertainties, are given in Table \ref{tbl-1}.

\begin{figure*}
\epsscale{1.15}
\plottwo{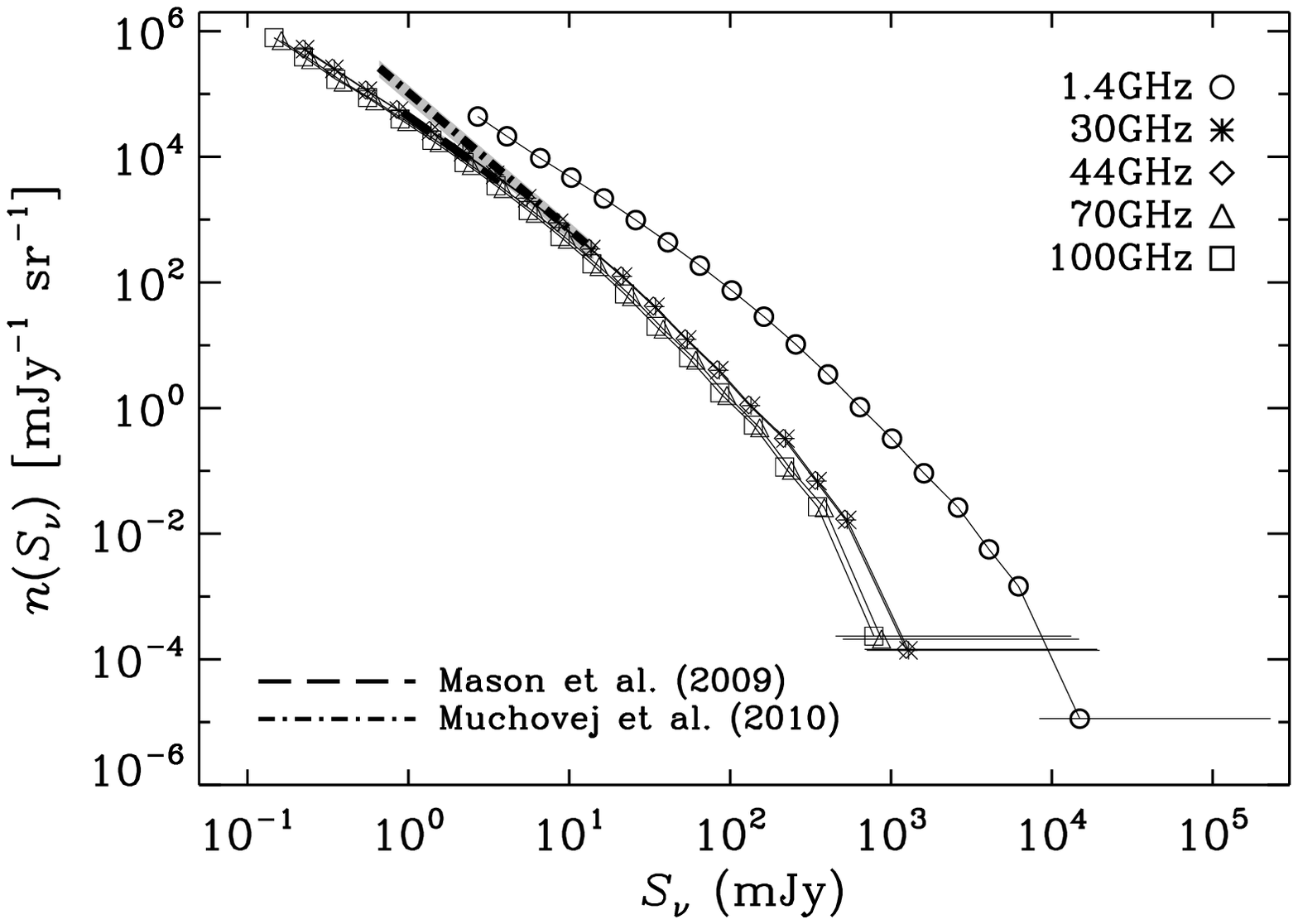}{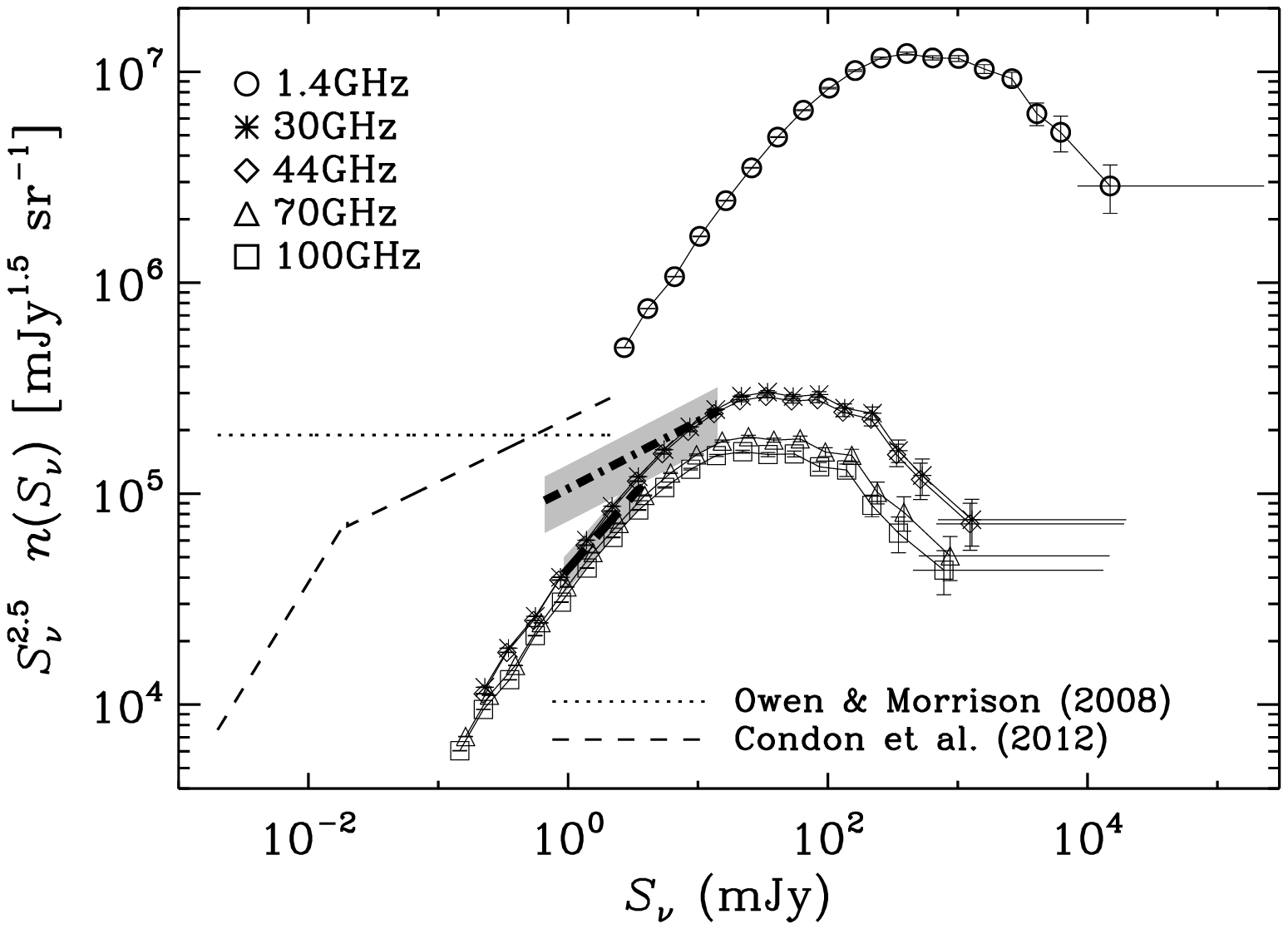}
\caption{{\it Left:} Differential source count estimates at 30, 44, 70, and 100\,GHz based on extrapolating the NVSS number counts combined with our stacking analysis.  
Error bars are not plotted as they are similar to the size of the plotting symbols.  
{\it Right:} Euclidean-normalized source counts at 30, 44, 70, and 100\,GHz based on NVSS detections combined with our stacking analysis of the {\it Planck} data.  
Similar as with the differential source counts in the left panel, the error bars are similar to the size of the plotting symbols, except for the brightest few flux density bins.
We additionally show the sub-mJy 1.4\,GHz source counts based on the results of \citet[][dotted line]{om08} and the recent $P(D)$ analysis of \citet[][dashed line]{jc12}.  
In both panels flux density bins have a width of 0.1\,dex except for the brightest flux density bins, whose widths are indicated by horizontal lines.  
The shaded gray region with the long-dashed line in both panels illustrates the predictions from \citet{bm09} for sources in the flux density range of  between $1 < S_{\rm 31\,GHz} < 4$\,mJy.  
We additionally show the predictions form \citet{sm10} (shaded regions with a dot-dashed line) for sources in the flux density range of  between $0.7 < S_{\rm 31\,GHz} < 15$\,mJy, which are discrepant with both our measurements and those from \citet{bm09}.    
}
\label{fig:dnds}
\end{figure*}

\section{Results}
In the following section we describe the results based on our stacking analysis.  
From these results, we are able to make estimates for the differential source counts and integrated extragalactic brightnesses from discrete sources at 30, 44, 70, and 100\,GHz.

\subsection{Spectral Indices}
Using the results of the stacked flux densities, we estimate the corresponding spectral index between each of the stacked {\it Planck} flux densities and the corresponding mean 1.4\,GHz flux density for that bin.  
Spectral indices are only calculated for 1.4\,GHz flux density bins that yielded a statistically significant (i.e., $>5\sigma$) stacked flux density at the {\it Planck} frequencies.       
The spectral indices are given in Table \ref{tbl-1} along with corresponding uncertainties and plotted in Figure \ref{fig:stkspx}.  
There is a clear trend of brighter 1.4 GHz sources having flatter spectral indices in all cases.  

Had we naively assumed that only the single, brightest NVSS source was associated with the location of our {\it Planck} cutouts, the total 1.4\,GHz flux densities would be reduced resulting in significantly flatter stacked spectral indices.  
This is especially true at 30\,GHz, where the beam area is $\approx$1900 times larger than that of NVSS.  
By including (weighted) contributions of nearby 1.4\,GHz sources, the associated 1.4\,GHz flux densities are larger by factors of $\approx$2.5, 2.1, 1.6, and 1.4 at 30, 44, 70, and 100\,GHz, respectively, compared to what is measured when only associating the cutout with the single, brightest NVSS source.  
This results in corresponding, stacked spectral indices that are steeper, on average, by factors of $\approx$1.3, 1.2, 1.1, and 1.1 at 30, 44, 70, and 100\,GHz, respectively; the differences are largest for the faintest flux density bins.  

To illustrate that our stacking analysis is in fact resulting in realistic spectral indices, we over plot the 1.4-to-30\,GHz spectral indices reported in \citet{crg15}.  
The spectral indices from our stacking analysis are clearly consistent with the mean and spread of values 
(asterisks) reported from direct detections of sources.  
Our values are also consistent with the results of \citet{bm09}, who reported on a 31\,GHz survey of 3165 known extragalactic radio sources included in NVSS using both the 100\,m Robert C. Byrd Green Bank Telescope and the 40\,m Owens Valley Radio Observatory telescope.
These authors measure a mean 1.4-to-31\,GHz special index of $-0.917$ with a standard deviation of $0.311$ among their entire sample; this is also illustrated in Figure \ref{fig:stkspx}.  
This is significantly steeper than the average 1.4-to-31\,GHz spectral index of $\sim -0.7$ reported by \citet{sm10}, which was measured from a sample of 209 sources detected at 31\,GHz with the Sunyaev-Zel'dovich Array and matched to 1.4\,GHz counterparts in NVSS.  
It is worth pointing out that any spectral index distribution having the same mean as that obtained from our stacking analysis will yield the same results.  
However, evolution in the spectral index distribution as a function of flux density within each bin is important and not directly measured, which is a limitation of our study.  

\subsection{Differential Source Counts}
Having the spectral indices per each 1.4\,GHz flux density bin, we use these to estimate the 30, 44, 70, and 100\,GHz flux densities for each NVSS source through a linear interpolation over stacked bins with statistically significant detections.  
This is done by first applying the stacked spectral index to all 1.4\,GHz sources that contributed to that specific stack to estimate a corresponding flux density for the {\it Planck} frequencies.    
For those NVSS sources having flux densities above or below where we were able to obtain a statistically significant detection through stacking, we use the spectral index from the highest and lowest 1.4\,GHz flux density bins, respectively.  
Then, an ordinary-least-squares fit to the scatter plots per {\it Planck} frequency were used to estimate the corresponding flux density from all sources in the NVSS catalog.  

Taking our estimated 30, 44, 70 and 100\,GHz flux densities, we generate corresponding differential source counts shown in the left panel of Figure \ref{fig:dnds}.  
In the right panel of Figure \ref{fig:dnds}, we additionally plot the Euclidian normalized source counts at all frequencies.  
In both panels, we compare with other differential source count estimates at $\sim$30\,GHz from the literature.  
The shaded region with the dot-dashed line illustrates the predicted 30\,GHz differential source counts over a flux density range of $0.7 < S_{\rm 31\,GHz} < 15$\,mJy reported by \citet{sm10}.  
The shaded region with the long-dashed line illustrates the predicted 30\,GHz differential source counts from \citet{bm09} over a flux density range of $1 < S_{\rm 31\,GHz} < 4$\,mJy.  
In both cases, to properly compare these measurements with ours at 28.5\,GHz, we have scaled their results using their average 1.4-to-31\,GHz spectral indices, which has a negligible effect.  

The differential source counts reported by \citet{bm09} are $\approx$16\% larger than what we calculate over the same flux density range from our stacking analysis, and thus in very good agreement.  
Such a difference could easily arise from selection biases and calibration errors in their study.  
The agreement between our source counts and \citet{bm09} may not be too surprising, as both populations are NVSS-selected.  
However, both the 30\,GHz differential source counts reported here and in \citet{bm09} are significantly lower than those reported by \citet{sm10}.  
These authors state that the discrepancy can be explained by a small shift in the spectral index distribution for faint 1.4\,GHz sources.  
However, even if we force a constant 1.4-to-30\,GHz spectral index of $-0.7$ for all stacked flux density bins, we are still  nowhere near able to increase our 30\,GHz differential source counts to match theirs.   
Furthermore, models for number counts of galaxies at these frequencies \citep[e.g.,][]{gdz05, mt11} appear to be much more consistent with what is reported here, and found by \citet{bm09}.  
In both cases, however, our methodology enables us to push the 30\,GHz source counts to an order of magnitude fainter in flux density,  
although we caution that our estimates at the faintest flux densities could be uncertain due to the absence of detections in the stacks at the faintest bins.    

To extend the differential source counts to flux densities below the 2.1\,mJy sensitivity limit of the NVSS catalog, we make use of recent work by \citet{jc12}, who employed a $P(D)$ analysis on a deep, confusion limited 3\,GHz observation of the Lockman Hole to estimate differential source counts down to a corresponding 1.4\,GHz flux density of $\approx$2\,$\mu$Jy.    
Following the conclusions of \citet{jc12}, we adopt functional forms for the differential 1.4\,GHz source counts of $n(S_{\rm 1.4\,GHz}) = 1.2\times10^{5}\,S_{\rm 1.4\,GHz}^{-1.5}\,{\rm Jy^{-1}\,sr^{-1}}$ \citep{jc84} and $n(S_{\rm 1.4\,GHz}) = 57\,S_{\rm 1.4\,GHz}^{-2.2}\,{\rm Jy^{-1}\,sr^{-1}}$ \citep{mc85} for sources in the flux density ranges of $2 < S_{\rm 1.4\,GHz} < 20\,\mu$Jy and $20 < S_{\rm 1.4\,GHz} < 2100\,\mu$Jy, respectively.  
These are shown in right panel of Figure \ref{fig:dnds} as a dashed line.  

As a comparison, we additionally make use of the Euclidean normalized differential source counts reported by \citet{om08}, which were also carried out over the Lockman hole using deep 1.4\,GHz observations [i.e., $n(S_{\rm 1.4\,GHz}) = 6\,S_{\rm 1.4\,GHz}^{-2.5}\,{\rm Jy^{-1}\,sr^{-1}}$].  
This is illustrated by a dotted line in Figure \ref{fig:dnds}
While the recent analysis of \citet{jc12} suggests that the source counts of \citet{om08} are likely too large, arising from an overcorrection of source brightnesses to integrated flux densities near the brightness cutoff of their catalog, we include these values in our analysis as they likely provide a conservative estimate for the uncertainty in our measurements.  

\begin{deluxetable}{cc|cccccc}
\tablecaption{Fit to Cumulative Source Counts \label{tbl-2}}
\tabletypesize{\scriptsize}
\tablewidth{0pt}
\tablehead{
\colhead{$\nu$} & \colhead{$>S_{\nu}$} & 
\colhead{} & \colhead{} & \colhead{} & \colhead{} & \colhead{} & \colhead{} \\ 
\colhead{(GHz)} & \colhead{($\mu$Jy)} & 
\colhead{$\xi_{0}$} & \colhead{$\xi_{1}$} & \colhead{$\xi_{2}$} & \colhead{$\xi_{3}$} & \colhead{$\xi_{4}$} & \colhead{$\xi_{5}$}
}
\startdata
   1&       2.0&     8.008&    0.6127&    -1.404&    0.4812&  -0.06781&  0.003233\\
  30&      0.17&     7.563&    -1.057&   -0.2775&    0.2245&  -0.04977&  0.003180\\
  44&      0.16&     7.548&    -1.063&   -0.2699&    0.2226&  -0.04976&  0.003199\\
  70&      0.11&     7.409&    -1.132&   -0.1773&    0.1916&  -0.04645&  0.003115\\
 100&      0.10&     7.357&    -1.146&   -0.1530&    0.1840&  -0.04593&  0.003129
\enddata
\end{deluxetable}

Finally, while not plotted, we additionally fit the cumulative source counts at each frequency with a fifth order polynomial such that,  
\begin{equation}
\label{eqn:ntgs}
\log\left[\frac{N(>S_{\nu})} {{\rm sr}^{-1}} \right]  = \sum_{i=0}^{5} \xi_{i} \log\left(\frac{S_{\nu}}{\rm \mu Jy}\right)^{i}~.
\end{equation}  
The corresponding coefficients from the fits, along with the faintest flux density used in the fit (i.e., $>S_{\nu}$), are given in Table~\ref{tbl-2}.   

\begin{deluxetable*}{c|ccccc}
\tablecaption{Integrated Extragalactic Light from Discrete Sources  \label{tbl-3}}
\tabletypesize{\scriptsize}
\tablewidth{0pt}
\tablehead{
\colhead{} &
\colhead{$T_{\rm b}$ (1.4\,GHz)} & \colhead{$T_{\rm b}$ (30\,GHz)} & \colhead{$T_{\rm b}$ (44\,GHz)} & \colhead{$T_{\rm b}$ (70\,GHz)} & \colhead{$T_{\rm b}$ (100\,GHz)}\\
\colhead{Component} &
\colhead{(mK)} & \colhead{($\mu$K)} & \colhead{($\mu$K)} & \colhead{($\mu$K)}  & \colhead{($\mu$K)}
}
\startdata
NVSS ($S_{\rm 1.4\,GHz}>2.1$\,mJy)&    68.68$\,\pm\,$     2.06&    14.12
$\,\pm\,$     0.44&     5.71$\,\pm\,$     0.17&  1.67$\,\pm\,$     0.05 &  0.74
$\,\pm\,$     0.02 \\
Owen \& Morrison (2008; $2 < S_{\rm 1.4\,GHz} < 2100\,\mu$Jy) &   135.83
$\,\pm\,$    20.35&    27.96$\,\pm\,$     5.97&    11.31$\,\pm\,$     1.83&
  3.32$\,\pm\,$     0.52 &  1.48$\,\pm\,$     0.20 \\
Condon et al. (2012; $2 < S_{\rm 1.4\,GHz} < 2100\,\mu$Jy) &    36.95$\,\pm\,$
    10.36&     7.64$\,\pm\,$     3.06&     3.09$\,\pm\,$     0.94&  0.92
$\,\pm\,$     0.27 &  0.41$\,\pm\,$     0.10 \\
\hline
Total [NVSS+\citet{om08}]&   204.52$\,\pm\,$    20.45&    42.08$\,\pm\,$
     5.99&    17.02$\,\pm\,$     1.84&  5.00$\,\pm\,$     0.52&  2.22$\,\pm\,$
     0.20\\
Total [NVSS+\citet{jc12}]&   105.63$\,\pm\,$    10.56&    21.76$\,\pm\,$
     3.09&     8.80$\,\pm\,$     0.95&  2.59$\,\pm\,$     0.27&  1.15$\,\pm\,$
     0.10
\enddata
\tablecomments{Temperatures are given in units of brightness temperature (see Section \ref{sec-stack}).  The measured central frequencies of the {\it Planck} data, which were used for the analysis, are 28.5, 44.1, 70.3, and 100\,GHz.}
\end{deluxetable*}

\subsection{The Integrated Extragalactic Light from Discrete Sources}
\begin{figure}
\epsscale{1.2}
\plotone{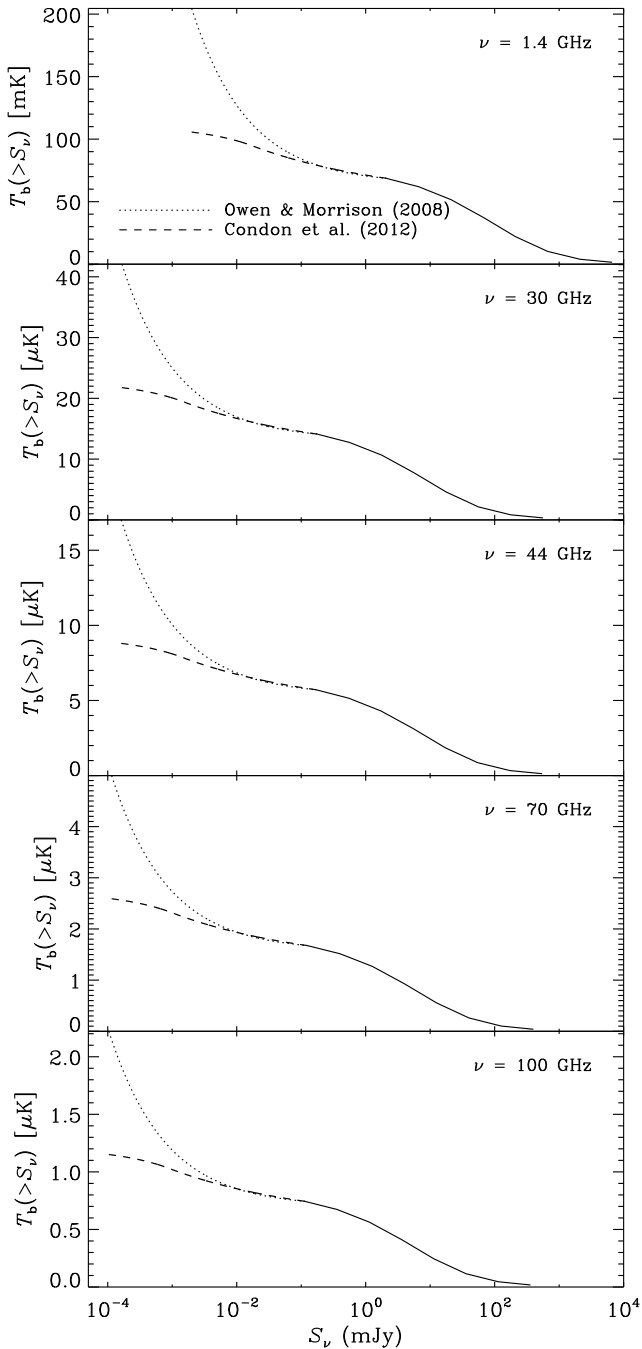}
\caption{Cumulative brightness temperature plots at 1.4, 30, 44, 70 and 100\,GHz.  
The solid line indicates measurements based on NVSS detected sources, whereas the dotted and dashed lines illustrate the behavior of the integrated light at sub-mJy flux densities based on the results of \citep[][]{om08} and \citep[][]{jc12}, respectively. }
\label{fig:cumTb}
\end{figure}

Cumulative plots of the integrated extragalactic light from individual sources at each frequency, in units of brightness temperature, are show in Figure \ref{fig:cumTb}.  
Extrapolations to the sub-mJy 1.4\,GHz population (i.e., $2 < S_{\rm 1.4\,GHz} < 2100\,\mu$Jy) based on the results of \citet{jc12} and \citet{om08} are also shown, clearly illustrating that the latter introduces a sharp (divergent) increase in the contribution from faint sources to the total extragalactic background light at these frequencies.  
In Table \ref{tbl-3}, the integrated brightness temperatures at 1.4, 30, 44, 70, and 100\,GHz for only NVSS sources (i.e.,  $S_{\rm 1.4\,GHz} > 2.1\,$mJy) are given.  
Also listed are the brightness temperature estimates for sources in the flux density range of $2 < S_{\rm 1.4\,GHz} < 2100\,\mu$Jy) based on the faint source count estimates from \citet{jc12} and \citet{om08}, where the latter results in a sky brightness that is nearly a factor of $\sim$2 larger than the former for $S_{\rm 1.4\,GHz} > 2\,\mu$Jy.  

While we only integrate down to $S_{\rm 1.4\,GHz} \approx 2\,\mu$Jy, which is the minimum flux density that the \citet{jc12} $P(D)$ analysis is sensitive to, continuing to integrate to lower flux densities results in the total brightness temperature from the \citet{om08} counts to increase rapidly, and eventually exceed the ARCADE\,2 total extragalactic sky brightness measurement.  
Specifically, by integrating down to 10\,nJy, we find that the integrated brightness temperature assuming the \citet{jc12} fit to the faint source population converges near 111\,mK, consistent within errors to what we obtain by only integrating down to where the $P(D)$ analysis is no longer sensitive.  
However, the corresponding brightness temperature assuming the \citet{om08} extrapolation to faint sources reaches a value of $\approx$63\,K, which clearly violates the ARCADE\,2 value, let alone the CMB temperature.  
This is of course due to the fact that to avoid Olbers' paradox, the differential source counts at faint flux densities must eventually achieve an index flatter than $-2.0$.

\begin{figure}
\epsscale{1.2}
\plotone{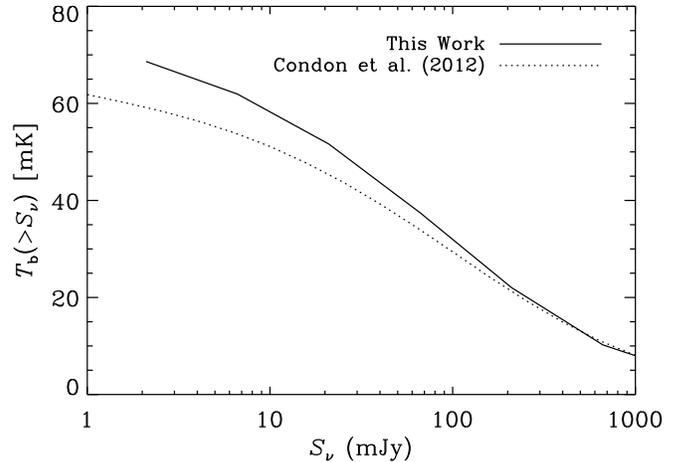}
\caption{A comparison of the cumulative 1.4\,GHz brightness temperature from this work with that of \citet{jc12} between 2.1\,mJy and 1\,Jy.  
There is a clear divergence occurring at $\approx$175\,mJy, where the integrated light from the NVSS data appears to be increase more rapidly towards lower flux densities compared to the curve from \citet{jc12}.  
This location of the divergence is exactly where the data used in the \citep{jc12} study switched from NRAO 91\,m data \citep{jm78} to various WSRT and VLA interferometric observations \citep[see][]{jc84} requiring large correction factors for incompleteness caused by primary beam attenuation, as well as for partial source resolution by the smaller synthesized beams.  }
\label{fig:cumTbcomp}
\end{figure}

\subsubsection{Comparison with Other Work at 1.4\,GHz}  
We note that the total contribution to the 1.4\,GHz brightness temperature from extragalactic sources reported here (i.e., $111\pm11\,$mK for $S_{\rm 1.4\,GHz} > 10\,$nJy) is slightly larger than that reported by \citet[][i.e., $100\pm10\,$mK for $S_{\rm 1.4\,GHz} > 10\,$nJy]{jc12} even when using the same estimates for the contribution from faint sources below the detection limit of the NVSS.  
To determine the cause of this discrepancy, we plot the cumulative 1.4\,GHz brightness temperature histograms using the data from \citet{jc12} and the NVSS data (i.e., this work) in Figure \ref{fig:cumTbcomp}. 
The cumulative histograms appear to diverge at $\approx$175\,mJy, where the 1.4\,GHz brightness temperature using the NVSS data begins to increase more rapidly towards lower flux densities relative to the histogram from \citet{jc12}.  
The location of the divergence is exactly where the data used in the \citet{jc12} study switched from single-dish, NRAO 91\,m data \citep{jm78} to various WSRT and VLA interferometric observations compiled by \citet{jc84}.  

Unlike the NVSS maps, which obtain nearly uniform coverage, these older interferometric data required large correction factors for incompleteness caused by primary beam attenuation, as well as for partial source resolution by the smaller synthesized beams. 
Had there been some sharp feature near the 2.1\,mJy limit of the NVSS survey data in Figure \ref{fig:cumTbcomp}, it would likely indicate that the NVSS is incomplete or unreliable near that limit.  
Additionally, if the NVSS correction for CLEAN bias was incorrect, it would likely lead to a conspicuous feature at the faint end.   
Given that we observe a smooth transition between the two curves in Figure \ref{fig:cumTbcomp} likely suggests that the discrepancy is most likely arising from underestimated corrections to the older VLA and WSRT data compiled in \citet{jc84}.  

\begin{figure*}
\epsscale{1.15}
\plottwo{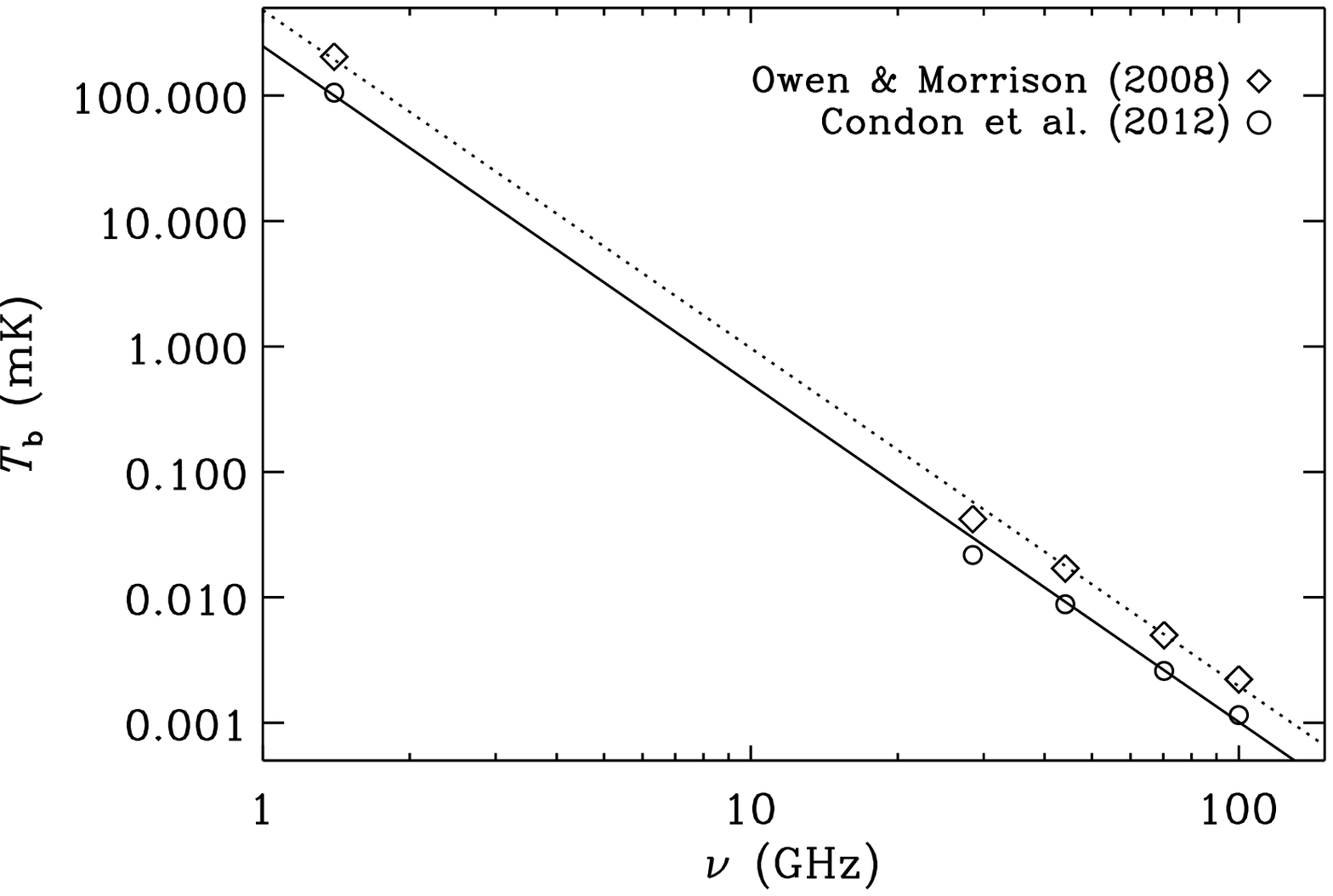}{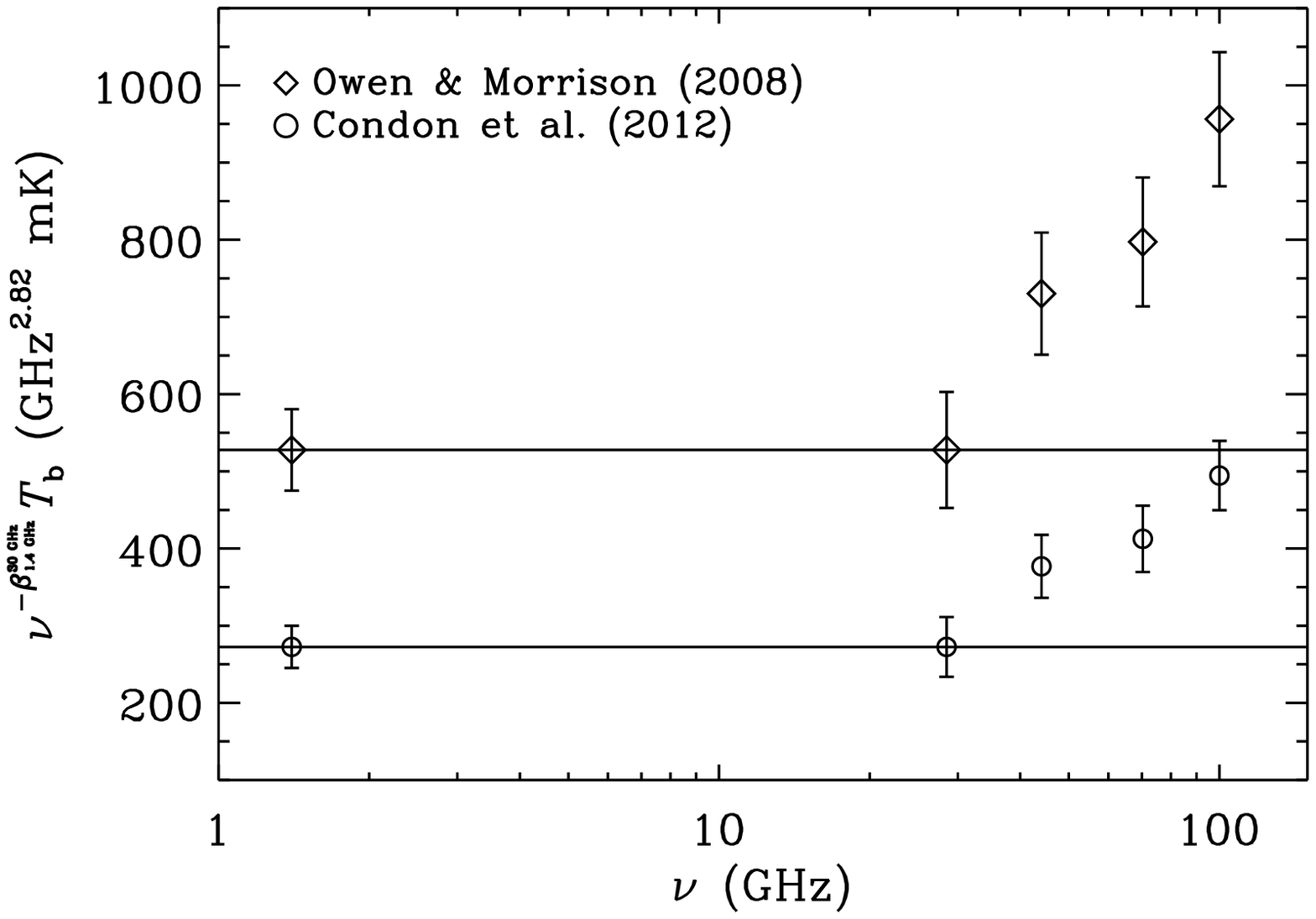}
\caption{{\it Left:} Extragalactic brightness temperature from discrete sources plotted against frequency using the differential source counts from both \citet{jc12} and  \citet{om08}.  
Power laws are fit to both sets of data, each having an index of $\beta_{\rm 1.4\,GHz}^{\rm 100\,GHz} = -2.69\pm0.03$.  
There appears to be an indication for spectral flattening at high frequencies, as power-law fits to only the {\it Planck} frequencies have an index of $\beta_{\rm 30\,GHz}^{\rm 100\,GHz} = -2.39\pm0.12$ in both cases.  
Error bars are not shown, as they are comparable to the plotting symbol size.  
{\it Right:} The same as plotted in the left panel, except that the brightness temperatures have been multiplied by $\nu^{-\beta_{\rm 1.4\,GHz}^{\rm 30\,GHz}}$, where $\beta_{\rm 1.4\,GHz}^{\rm 30\,GHz} = -2.82\pm0.06$.  
In doing this, departures from a single power law fit, identified by the horizontal lines and indicating spectral flattening of the integrated brightness temperatures with increasing frequency, can easily be shown to be statistically significant (i.e., $6.5\sigma$) given the associated error bars on the individual brightness temperature measurements.  
}
\label{fig:Tb}
\end{figure*}

\begin{figure}
\epsscale{1.2}
\plotone{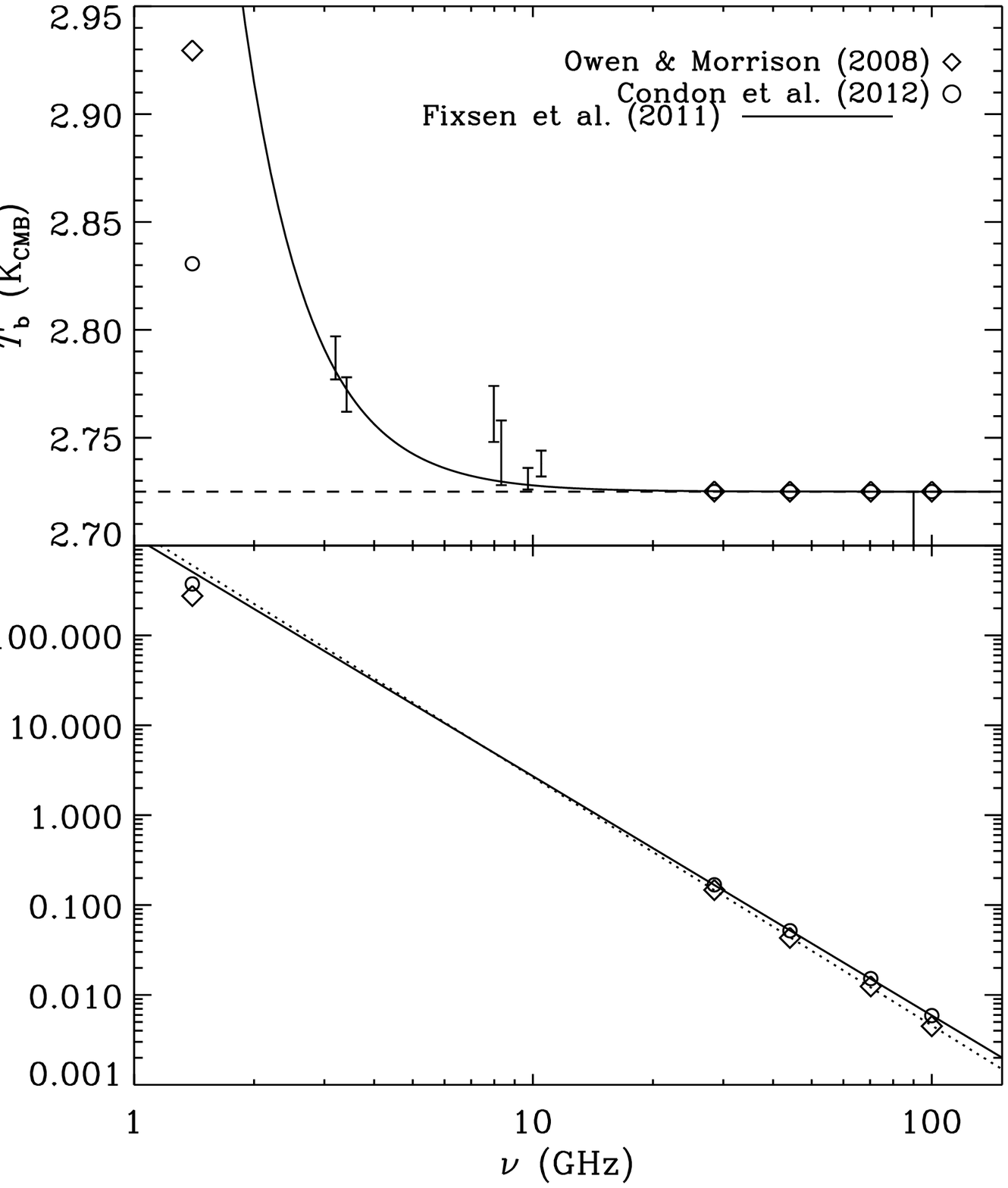}
\caption{{\it Top:} Thermodynamic temperature (in units of ${\rm K_{CMB}}$) compared to \citet[][solid line]{djf11}, which is a best fit to the ARCADE\,2 data modeled by a constant CMB temperature plus a synchrotron-like component having an index of $-2.6$.  
The dashed line indicates the 2.725\,K temperature of the CMB.  
Even using the much larger source count estimates from \citet{om08}, the total integrated 1.4\,GHz extragalactic light is still significantly less than the fit to the ARCADE\,2 measurement.
{\it Bottom:}  The excess temperature between the ARCADE\,2 fit and our integrated extragalactic brightness temperatures at 1.4, 30, 44, 70 and 100\,GHz assuming the sub-mJy number counts reported by both \citet{jc12} and \citet{om08}.  
The power-law fits to the excesses have indices of $\beta = -2.66$ and $-2.76$, respectively.  
}
\label{fig:arccomp}
\end{figure}

\subsection{Comparison with the ARCADE\,2 Excess}
In the left panel of Figure \ref{fig:Tb} the integrated brightness temperatures from discrete sources at 1.4, 30, 44, 70 and 100\,GHz are plotted against frequency using the faint source contribution estimates from both \citet{jc12} and \citet{om08}.  
These data are marginally well fit by a single power law having an index of $\beta_{\rm 1.4\,GHz}^{\rm 100\,GHz} = -2.69\pm0.03$.  
If we instead only fit the data at the {\it Planck} frequencies, we obtain an index that is significantly flatter, being $\beta_{\rm 30\,GHz}^{\rm 100\,GHz} = -2.39\pm0.12$.  
This is illustrated in the right panel of Figure \ref{fig:Tb}, where the integrated brightness temperatures have been multiplied by $\nu^{-\beta_{\rm 1.4\,GHz}^{\rm 30\,GHz}}$.  
These values are consistent using either \citet{jc12} or \citet{om08} for the faint source count estimates, being $\beta_{\rm 1.4\,GHz}^{\rm 30\,GHz} = -2.82\pm0.06$.  
This is a statistically significant (i.e., 6.5$\sigma$) indication that the extragalactic sky brightness spectrum from discrete sources is flattening with increasing frequency.  

To ensure that this result is not driven by the extremely steep 1.4-to-30\,GHz spectral index measured for the faintest 1.4\,GHz flux density bin, we determine how much this value would have to change (i.e., flatten) to reduce the significance of the measured excess at 44, 70, and 100\,GHz to $<3\sigma$.  
For this to occur, we find that the 1.4-to-30\,GHz spectral index would need to flatten from $-0.89$ to a value that is $\gtrsim -0.77$, a change of $\gtrsim 4 \sigma$.  
This would suggest that our 1.4\,GHz flux densities are being grossly overestimated for sources in that flux density bin, which seems unlikely.  
We additionally make sure that this result is not being significantly affected by potential Galactic CO emission that could be contaminating our 100\,GHz flux densities. 
We therefore run the entire analysis after reducing the stacked 100\,GHz flux densities by 4.5\%, which is the median excess contribution of the measured 100\,GHz flux densities of sources in the {\it Planck} Early Release Compact Source Catalog \citep[ERCSC][]{chen16}.  
We find that this has little effect, reducing the significance of the 44, 70, and 100\,GHz excesss found in the right panel Figure \ref{fig:Tb} from 6.5 to 6.1$\sigma$.  

In the top panel of Figure \ref{fig:arccomp}, thermodynamic temperatures from our integrated extragalactic brightness temperatures, added to a CMB temperature of $2.725\pm0.001$\,K \citep{djf11}, are plotted against frequency again using the faint source contribution estimates from both \citet{jc12} and \citet{om08}.  
As recently pointed out by \citet{jc12}, the integrated light from individual sources at GHz frequencies is substantially less than the fit to the ARCADE\,2 extragalactic background measurements \citep{djf11}, which is given by the solid line [$T_{\rm b} = (24.1 \pm 2.1\,{\rm K_{CMB}})(\nu/{\rm 310\,MHz})^{(-2.599\pm0.036)}$] and the points with error bars.  
We find this to be true even if we assume the \citet{om08} sub-mJy differential source count results and only integrating down to a 1.4\,GHz flux density of 2\,$\mu$Jy.  
Of course, as stated above, by integrating down to fainter flux densities and assuming that the shape of the differential sources counts reported by \citet{om08} persists, results in integrated brightness temperates that greatly exceeds the limits imposed by ARCADE\,2.  

In the bottom panel of Figure \ref{fig:arccomp}, the residuals between the fit to the ARCADE\,2 data from \citet{djf11} and our integrated thermodynamic temperatures are plotted against frequency using the integrated source contribution estimates from \citet{jc12} and \citet{om08}.  
Each of these results in positive values.  
After removing the CMB temperature, the fit to the ARCADE\,2 extragalactic sky brightness measurements is a factor of $\approx$4.5, 8.6, 6.6, 6.2, and 4.9 times brighter than what we measure from discrete sources at 1.4, 30, 44, 70, and 100\,GHz, respectively, assuming the \citet{jc12} differential source counts at sub-mJy flux densities.  
If we instead assume the differential source counts of \citet{om08}, the ARCADE\,2 total extragalactic sky brightness is still a factor of $\approx$2.3, 4.4, 3.4, 3.2, and 2.5 times brighter than what we measure from discrete sources at 1.4, 30, 44, 70, and 100\,GHz.  
The above comparison between our measurements of the sky brightness from discrete sources at the {\it Planck} frequencies to what is estimated from the fit to the ARCADE\,2 data may not be too enlightening, given that their fit was not constrained by data at frequencies $\gtrsim$10\,GHz.  
However, we note it here as it does yield values that are broadly
consistent to the excess measured at $\sim$1\,GHz. 
Furthermore, the spectrum of the residual is largely consistent with diffuse Galactic synchrotron emission.  


\begin{figure}
\epsscale{1.2}
\plotone{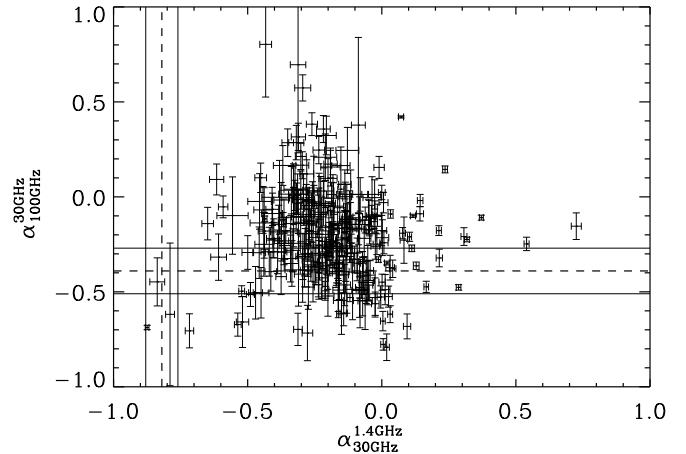}
\caption{The 1.4-to-30\,GHz spectral index plotted against the spectral index measured across the 4 {\it Planck} frequencies spanning 30--100\,GHz for detections (i.e., $\vert b \vert  > 20\degr$) in the band-merged {\it Planck} ERCSC \citep[][]{chen16}.  
As with the stacking analysis, the 1.4\,GHz flux density used in calculating the spectral indices is the
weighted sum of all NVSS sources to simulate what would be measured for the much larger corresponding {\it Planck} beam.  
The corresponding spectral indices for the integrated discrete-source extragalactic spectrum, along with 1$\sigma$ error bars, are shown.  
The minimum 30\,GHz flux density for sources included in this plot is $\approx$387\,mJy, such that only 1.4\,GHz sources having flux densities larger than $\approx$4.6\,Jy would have a spectral index consistent with what is measured for the integrated light of discrete sources (i.e., a negligible fraction of NVSS sources). 
We note that there is one source in the sample whose spectrum appears consistent with that of the integrated light in the vicinity of NGC\,5832, but does not appear to be unassociated with the galaxy.  
}

\label{fig:ercsccomp}
\end{figure}

\section{Discussion} 

We have used a stacking analysis to estimate source counts at 30, 44, 70 and 100\,GHz.  
In doing this, we have also been able to estimate the total contribution from discrete sources to the extragalactic sky background at 30, 44, 70 and 100\,GHz.  
These results
One of the unexpected results from this analysis is that the discrete-source extragalactic sky brightness temperature appears to flatten with increasing frequency, as shown in the right panel of Figure \ref{fig:Tb}. 

We fit the measured broadband spectrum of the radio background as the sum of two power-laws, as this would be the functional form for high-frequency synchrotron emission.  
The spectral index of the steep component is fixed to $-2.85$, as this is the minimum value for which fitting two components yields a smooth spectrum, and sets an upper limit for the contribution of the high-frequency component.  
The resultant spectral energy distribution of the discrete-source extragalactic brightness temperature is of the form T$_{b}=275.6\times\nu^{-2.85}+2.31\times\nu^{-1.79}$.
If we had instead adopted the NVSS (S$_{\rm 1.4\,GHz}>2.1$\,mJy) limit, we find little change to the spectral indices but find that the coefficients are naturally reduced i.e. T$_{b}=179.2\times\nu^{-2.85}+1.55\times\nu^{-1.80}$.
If this is synchrotron emission, it implies that there is a component of high energy particles with a harder spectrum than what produces the synchrotron emission at low frequencies.
We find that between 30 and 100\,GHz the harder component accounts for $\gtrsim$30\% of the total brightness temperature relative to what is measured when extrapolating the lower frequency  component between 1.4 and 30\,GHz to the higher {\it Planck} frequencies.  
For an electron injection spectrum of $E^{-p}$, the intensity of optically-thin synchrotron emission has a spectral index of $\alpha=-(p-1)/2$. Since $\alpha=2+\beta$,
the injected spectrum of particles that are responsible for the high frequency emission must be extremely hard $p=0.2$. 

One possibility is that a fraction of Gigahertz-peaked radio sources (GPS), which have synchrotron self-absorption turnovers between $40-100$\,GHz, are contributing to the upturn since these sources would be several orders of magnitude fainter than the NVSS detection limit at 1.4\,GHz. In the band-merged {\it Planck} ERCSC \citep{chen16}, among sources detected at all {\it Planck} frequencies, we found that 43\% of radio sources peak at 44\,GHz, with 10\% peaking at 70\,GHz and 5\% at 100\,GHz, which alone are not able to account for the measured flattening of the extragalactic background light from discrete sources.  
We find that to account for the observed flattening of the spectral index, there must be a higher fraction of GPS sources with increasing frequency. 
That is, for the sample of sources observed at 1.4\,GHz with a spectral index of $-0.82$, we would need $\approx 20$\% more sources peaking at 44\,GHz and $\approx 25$\% more sources peaking at 70 and 100\,GHz to account for the spectrum of the extragalactic background light from discrete sources.
There is definitely evidence that such a population of GPS sources exist at the higher frequencies although whether the fractions are as large as this remains unclear \citep{Planck2016-XLV}

Alternately, to account for the spectrum of the extragalactic background light from discrete sources., the average typical spectrum of the sources must evolve from $\alpha=-0.82$ between 1.4 and 30\,GHz to $\alpha=-0.39$ between 30 and 100\,GHz as shown by the sources in the bottom left corner of Figure \ref{fig:ercsccomp}. 
The origin for such a hard spectrum of energetic electrons is unclear. 
High frequency observations of individual sources in multiple bands may shed more light but we discuss possible origins for such a hard component of emission below.

\subsection{Frequency Selection?}
Looking at a $10\degr\times10\degr$ patch of the S$^{3}$ simulated extragalactic sky \citep{s-cubed08}\footnote{The simulation database can be accessed online via http://s-cubed.physics.ox.ac.uk.}, $\approx$75\% of sources above the NVSS surface brightness limit are classified as FR-I radio galaxies \citep{fr74}, for which their brightness decreases from the central galaxy.  
The more luminous FR-II radio galaxies, which are identified by hot spots in their lobes at a large distance from the central galaxy core, make up $\approx$20\% of sources above the NVSS surface brightness limit. 
Given the non-uniform magnetic fields and electron energy distributions in these geometrically complicated structures, along with the complexity of acceleration physics, their spectra exhibit a range in behavior and can deviate from a single, simple power law.    

The observed spectrum, therefore, might simply be the result of combining different classes of radio galaxies.  
For instance, looking at detections in the band-merged {\it Planck} ERCSC \citep[][]{chen16}, we can identify a handful of sources that have spectra similar to what we observe for the integrated extragalactic discrete-source spectrum (Figure \ref{fig:ercsccomp}).  
The fact that only a handful of sources have similar spectra may not be that surprising given that the faintest 30\,GHz flux density is $\approx$387\,mJy, requiring a 1.4\,GHz of $\approx$4.6\,Jy (i.e., a negligible fraction of NVSS sources) to obtain a 1.4-to-30\,GHz spectral spectral index of $-0.82$.  


\begin{figure}
\epsscale{1.2}
\plotone{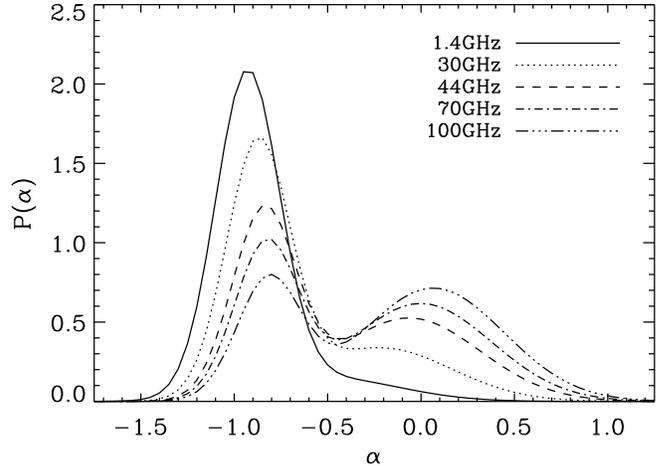}
\caption{The probability distribution (integral unity) of spectral indices calculated as a function frequency of based on the simple two component model presented in \citet{jc84}.    
The model consists of a steep and flat spectral index distribution described by Gaussians based on 1.4 and 5\,GHz observations.  
The average spectral indices flatten with increasing frequency as the flat spectrum component begins to dominate, taking mean values of $-0.87, -0.45, -0.34, -0.22,$ and $-0.12$ at 1.4, 30, 44, 70, and 100\,GHz,  respectively.  
}
\label{fig:spxdist}
\end{figure}

In Figure \ref{fig:spxdist}, we show probability distributions for spectral indices calculated at each of the frequencies being investigated here based on a simple two component model presented in \citet{jc84}.  
The observed spectral index distribution consists of a population of steep and flat power-law spectrum sources with a differential source count slope of $-2$ modeled by two Gaussians  \citep[see Appendix of][]{jc84}.  
At the fiducial frequency of 1.4\,GHz the steep component is defined by a Gaussian with amplitude 0.86, mean $-0.93$, and standard deviation $0.17$ at $z=0$ while the flat component is defined by a Gaussian with amplitude 0.14, mean $-0.50$, and standard deviation $0.38$.  
Since the spectral index distributions have a finite width, as one selects sources at higher frequencies the observed spectral index will naturally flatten even if just considering the steep spectral index population alone.  
Furthermore, as one selects sources at higher frequencies, the contribution of the flat spectrum population begins to increase, leading to even flatter average spectral indices, with a mean shifting between $-0.87$ at 1.4\,GHz and $-0.12$ at 100\,GHz for this simple model.    

While the mean spectral index measured at the middle of the {\it Planck} frequencies is $-0.26$, $\approx 1\sigma$ flatter than what is measured here, this may be expected since flat spectrum sources selected at 1.4 and 5\,GHz arising from synchrotron self-absorption will in fact steepen at higher frequencies as they become optically thin. 
Thus, the distribution of spectral indices arising from the combination of such sources and GPS sources could certainly lead to the observed discrete-source extragalactic spectrum, which may be the most plausible explanation.

\subsection{Free-Free Emission?}
Another explanation for the spectral flattening could be from free-free emission associated with star formation.
If we instead fit the discrete-source extragalactic spectrum with a combination of a synchrotron component, having a spectral index equal to that measured between 1.4 and 30\,GHz (i.e., $\alpha = -0.82$, which would actually be an upper limit on the non-thermal spectral index), along with a thermal (free-free) emission component having a spectral index of $\alpha = -0.1$ requires a thermal fraction at 30\,GHz of $\gtrsim$50\%.  
This is consistent to what is found for typical star-forming galaxies in the universe \citep[e.g.,][]{cy90,jc02}.  
However, over half of the extragalactic background contribution from discrete sources at 1.4\,GHz is accounted for by sources stronger than $S_{\rm 1.4\,GHz} \gtrsim 2.1\,$mJy.  
At these flux density levels, sources are typically AGN \citep[e.g.,][]{jc84, jc02, s-cubed08}.  

For instance, again looking at a $10\degr\times10\degr$ patch of the S$^{3}$ simulated extragalactic sky, indicates that only $\approx$1\% of sources above the NVSS surface brightness limit are star-forming galaxies.    
Consequently, we do not necessarily expect to see a spectral flattening at $\gtrsim$30\,GHz that one finds in star-forming galaxies arising from free-free emission, unless a significant fraction of such AGN also host ongoing star formation that is not captured in such simulations.   
However, spectra of canonical radio galaxies appear synchrotron-dominated through 100\,GHz, rather than having an M\,82-like spectrum that becomes free-free dominated beyond $\sim$30\,GHz.  
While difficult to assess given the data in-hand, this explanation for the flattening of the discrete-source extragalactic sky brightness temperature by free-free emission associated with star formation seems unlikely.

\begin{figure}
\epsscale{1.1}
\plotone{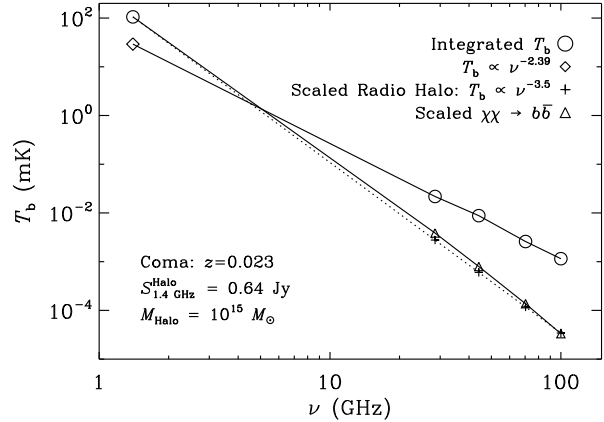}
\caption{  
The open circles show the measured discrete-source extragalactic sky brightness temperature measured.  
The plus symbols (dotted-line) are used to illustrate a scaled radio halo spectrum having a spectral index of $\beta = -3.5$, similar to that of the Coma cluster halo, which has a total 1.4\,GHz flux density of 0.64\,Jy \citep{deiss97}.   
The triangles show the DM annihilation spectrum for $\chi\chi \rightarrow b\bar{b}$ scaled to match the measured sky brightness from discrete sources at 1.4\,GHz.  
The DM annihilation spectrum is based on the models \citet{estorm16} for a Coma-like cluster at $z=0.023$ having a halo mass of $10^{15}\,M_{\sun}$, DM mass of $m_{\chi} = 100$\,GeV, and cross section $\langle \sigma v \rangle = 3 \times 10^{-26}\,$cm$^{3}$\,s$^{-1}$.  
The diamond plots the 1.4\,GHz brightness temperature extrapolated from the values measured at the four {\it Planck} bands that are fit with a spectral index of $\beta = -2.39$.  
The difference between this value and the measured sky brightness temperature at 1.4\,GHz is 76.4\,mK, which in turn would correspond to a contribution from DM annihilation for $\approx 7.4\times10^{6}$ or radio halo emission for $\approx 4.4\times10^{5}$ Coma-like clusters to account for a synchrotron excess at the lower frequencies if indeed the extragalactic sky brightness temperature from galaxies was represented by a single power law having a spectral index of $\beta = -2.39$.  
}
\label{fig:dm}
\end{figure}

\subsection{Positron Excess?}
It is worth seeing if the observed spectral flattening in the discrete-source extragalactic sky brightness temperature could be related to the excess of positrons reported by both the Payload for Antimatter-Matter Exploration and Light nuclei Astrophysics \citep[PAMELA;][]{pamela09} experiment and Alpha Magnetic Spectrometer on the International Space Station \citep[AMS-02;][]{ams13,ams14}.  
If we are to assume a typical magnetic field strength of $\sim$5\,$\mu$G in the host of radio galaxies, the energies for which cosmic-ray electrons emit at a critical frequency corresponding to the 4 {\it Planck} frequencies are 21, 26, 33, and 39\,GeV.  
Integrating over these energies, PAMELA and AMS-02 report a fractional excess of positrons 
that corresponds to in an increase of $\approx$8\% to the sky brightness temperature.  
This is well below the $\approx$30\% excess in brightness temperature integrated over those same energies compared to what one would expect if the brightness temperature spectrum had a spectral index of $\beta = -2.82$ between 1.4 and 100\,GHz, suggesting that the two are unrelated.  

\subsection{Radio Halos?}
Another possible explanation for the observed spectral flattening in the discrete-source extragalactic sky brightness temperature could be related to additional emission from steep-spectrum (i.e., $\alpha < -1$) radio halos associated with the massive clusters that many of the brightest radio sources that contribute significantly to the extragalactic sky brightness reside in.  
In this case, the assumption would be that the spectra from individual galaxies is in fact much flatter, similar to what is measured between 1.4 and 100\,GHz for the ERCSC sources plotted in Figure \ref{fig:ercsccomp} (i.e., $\beta=-2.27$ between 1.4 and 100\,GHz), and there is a significant contribution of synchrotron emission from radio halos that is steepening the low-frequency end of the spectrum.  

In Figure \ref{fig:dm} we plot a model radio halo synchrotron spectrum having a spectral index of $\beta = -3.5$ \citep{feretti12} scaled to the measured sky brightness temperature from discrete sources at 1.4\,GHz.  
The 1.4\,GHz flux density of the radio halo in Coma is $S_{\rm 1.4\,GHz} = 0.64\pm0.035$\,Jy \citep{deiss97}.  
Assuming that the sky brightness temperature from discrete sources indeed has a flat spectrum consistent with what is measured across the 4 {\it Planck} bands (i.e., $\beta = 2.39$), the corresponding 1.4\,GHz brightness temperature would be 29.21\,mK, which is 76.42\,mK less than what is measured.  
Consequently, it would take a contribution from $\approx 4.4\times10^{5}$ such radio halos (i.e, $\approx$37\% of all NVSS sources considered here) to account for this difference in the integrated sky brightness temperature, and thus is an unsatisfactory explanation.

\subsection{Dark Matter Annihilation?}
A final, and far more speculative, explanation for the observed spectral flattening in the discrete-source extragalactic sky brightness temperature could be related to the additional emission from dark matter (DM) annihilation that is expected to be associated with the massive clusters.  
Similar to a potential explanation by radio halos, this scenario assumes that the observed sky brightness temperature from galaxies is indeed quite flat but there is some excess component that is steepening what is observed at the low frequency end.  

In Figure \ref{fig:dm} we investigate such a scenario by looking at the possible contribution to the observed synchrotron spectrum from DM annihilation for a Coma-like cluster at $z=0.023$ having a halo mass of 10$^{15}\,M_{\sun}$ using the models of \citet{estorm16}.   
We consider DM annihilation to both $b\bar{b}$ and $\mu^{+}\mu^{-}$ integrated over a circular area having a radius of 300\,kpc (20\arcmin), assuming a DM mass of $m_{\chi} = 100$\,GeV and cross section $\langle \sigma v \rangle = 3 \times10^{-26}\,$cm$^{3}$\,s$^{-1}$.  
The spectrum from DM annihilation to $\mu^{+}\mu^{-}$ has a spectral index of $\beta = -2.65$ between 1.4 and 100\,GHz, 
$\approx 3 \sigma$ flatter than the spectral index of $\beta=-2.82\pm0.06$ for the measured sky brightness temperature from discrete sources between 1.4 and 30\,GHz, and thus may be excluded out as a candidate explanation.  
However, the spectrum from DM annihilation to $b\bar{b}$ has a spectral index of $\beta = -3.26$ between 1.4 and 100\,GHz, suggesting that it could potentially contribute to any spectral steepening at the low frequency end of our sky brightness temperature measurements.  
In Figure \ref{fig:dm} we scale this DM annihilation spectrum to the measured sky brightness temperature at 1.4\,GHz in Figure \ref{fig:dm} as triangles.  

Again, assuming that the sky brightness temperature from discrete sources indeed has a flat spectrum consistent with what is measured across the 4 {\it Planck} bands and that this difference is associated solely attributed to the contribution from DM annihilation for Coma-like halos based on the above assumptions and models, it would take $\approx7.4\times10^{6}$ such massive clusters, which is orders of magnitude beyond reality.    
If the assumed cross section is incorrect, and a value an order of magnitude smaller is more appropriate (i.e., $\langle \sigma v \rangle = 3 \times10^{-27}\,$cm$^{3}$\,s$^{-1}$), the situation only becomes worse as it would in turn require an order or magnitude more clusters.  
Consequently, an explanation from DM annihilation seems highly unlikely.  

\section{Conclusions}

By stacking {\it Planck} all sky maps at the location of NVSS sources, we are able to make estimates for the differential source counts and extragalactic brightness temperature for discrete sources at 30, 44, 70 and 100\,GHz.  
Our conclusions can be summarized as follows:
\begin{itemize}

\item{Assuming that the differential source counts for sources at 1.4\,GHz have been accurately estimated down to $S_{\rm 1.4\,GHz} \gtrsim 2\,\mu$Jy by \citet{jc12}, we measure integrated extragalactic sky brightnesses from discrete sources of $105.63 \pm 10.56\,$mK, $21.76\pm 3.09\,\mu$K, $8.80 \pm 0.95\,\mu$K,  $2.59\pm 0.27\,\mu$K, and $1.15\pm 0.10\,\mu$K at 1.4, 30, 44, 70, and 100\,GHz, respectively.  
After removing the CMB temperature, the fit to the ARCADE\,2 total extragalactic sky brightness measurements is a factor of $\approx$4.5, 8.6, 6.6, 6.2, and 4.9 times brighter than what we measure from discrete sources at 1.4, 30, 44, 70, and 100\,GHz, respectively. }

\item{We find evidence for the extragalactic radio spectrum from discrete sources to flatten with increasing frequency, having a spectral index of $\beta = -2.82\pm0.06$ between 1.4 and 30\,GHz, flattening to $\beta -2.39\pm0.12$ between 30 and 100\,GHz. 
This corresponds to a integrated radio spectrum (i.e, $S_{\nu} \propto \nu^{\alpha}$) from extragalactic sources whose spectral index flattens from $\alpha = -0.82\pm0.06$ between 1.4 and 30\,GHz to $\alpha = -0.39\pm0.12$ between 30 and 100\,GHz.  
We believe that the spectral flattening most likely arises from the sheer complexity of radio galaxy spectra that results in a range of spectral indices; 
in particular, there must be a set of sources with a harder spectrum at frequencies $>$30\,GHz although the origin of this energetic component of electrons is unclear.      
}

\item{
By integrating down to 10\,nJy, we obtain a slightly larger integrated 1.4\,GHz extragalactic brightness temperature (i.e., $111 \pm 11\,$mK) than that reported by \citet[][i.e., $100\pm10\,$mK]{jc12}.  
This difference most likely arises from having better estimates for the flux densities of detected sources using NVSS data rather than the older interferometric data compiled by \citet{jc84}.  
}


\end{itemize}

\acknowledgements
We would like to thank the anonymous referee for very useful comments that helped to improve the content and presentation of this paper. 
E.J.M. would like to thank J.J. Condon for many useful discussions that helped improve the presentation of the paper.  
E.J.M. would also like to thank B. Rusholme for providing modified software that significantly sped up the analysis of this investigation, E. Storm for providing her DM annihilation spectra for this study, and B. Mason and C. Sarazin for helpful discussions on cluster radio emission.  
The results of this paper are based on observations obtained with {\it Planck}, an ESA science mission with instruments and contributions directly funded by ESA Member States, NASA, and Canada.
The National Radio Astronomy Observatory is a facility of the National Science Foundation operated under cooperative agreement by Associated Universities, Inc.
This research has made use of the NASA/IPAC Extragalactic Database (NED), as well as the NASA/ IPAC Infrared Science Archive, both of which are operated by the Jet Propulsion Laboratory, California Institute of Technology, under contract with the National Aeronautics and Space Administration.
This research has made use of the VizieR catalogue access tool, CDS, Strasbourg, France.  

\bibliography{aph.bbl}

\end{document}